\documentclass[11pt,a4paper]{article}
\pdfoutput=1

\usepackage{jheppub}
\usepackage{amsmath}
\usepackage{bbm}
\usepackage{color}
\usepackage{placeins}

\graphicspath{ {fig/} }

\allowdisplaybreaks[4]

\DeclareMathOperator{\Li}{Li}

\newcommand{\be}{\begin{equation}}
\newcommand{\ee}{\end{equation}}
\newcommand{\bea}{\begin{eqnarray}}
\newcommand{\eea}{\end{eqnarray}}
\newcommand{\nn}{\nonumber}
\newcommand{\gpls}{GHPLs}
\newcommand{\gpl}{GHPL}
\newcommand{\Xxi}{c_\theta^2}

\makeatletter
\renewcommand\@fpheader{\hfill \parbox{3cm}{BI-TP~2013/21\\MITP/13-051\\ROMA1-1473/13\\ZU-TH~20/13}}
\renewcommand\@journal{}
\makeatother

\title{Light-quark two-loop corrections to heavy-quark pair production
in the gluon fusion channel}

\author[a]{R.~Bonciani\footnote{On leave of absence from Dipartimento di Fisica,
Universit\`a di Roma ``La Sapienza''.},}
\author[b]{A.~Ferroglia,}
\author[c]{T.~Gehrmann,}
\author[dc]{A.~von~Manteuffel,}
\author[e]{C.~Studerus}

\affiliation[a]{
  INFN Sezione di Roma,
  Piazzale Aldo Moro 5,
  00185 Roma, Italy
}
\affiliation[b]{
  Physics Department, 
  New York City College of Technology,
  The City University of New York, 
  300 Jay Street Brooklyn,
  NY 11201  US
}
\affiliation[c]{
  Institute for Theoretical Physics,
  University of Z\"urich,
  Winterthurerstrasse 190,
  8057 Z\"urich, Switzerland
}
\affiliation[d]{
  PRISMA Cluster of Excellence \& Institute of Physics,
  Johannes Gutenberg University,
  55099 Mainz, Germany
}
\affiliation[e]{
  Faculty of Physics,
  University of Bielefeld,
  Postfach 100131,
  33501 Bielefeld, Germany
}

\emailAdd{roberto.bonciani@roma1.infn.it}
\emailAdd{aferroglia@citytech.cuny.edu}
\emailAdd{thomas.gehrmann@uzh.ch}
\emailAdd{manteuffel@uni-mainz.de}
\emailAdd{cedricstuderus@gmail.com}

\abstract{
We calculate the two-loop corrections to heavy-quark pair production in the
gluon fusion channel which arise from diagrams involving a closed light-quark
loop. The calculation is carried out by keeping the exact dependence on the
heavy-quark mass. The analytic results are written in terms of logarithms,
classical polylogarithms $\Li_n$ ($n=2,3,4$), and genuine multiple
polylogarithms $\Li_{2,2}$. The functional arguments are rational expressions of
two independent external invariants and they are chosen in such a way that the
functions are real in all the physical phase-space points. Through systematic
changes in the functional basis, we obtain expansions of the results in both the
production threshold and small mass limits.}

\begin{document}

\maketitle


\section{Introduction}
\label{sec:intro}


The production of top-antitop quark pairs is one of the processes which  is
measured with such a precision at the Large Hadron Collider (LHC) that, on the
theoretical side, one needs to obtain predictions including perturbative
corrections beyond the  next-to-leading order (NLO) in QCD. Recently, the 
calculation of the full next-to-next-to leading order (NNLO)
corrections to the total top-quark pair-production cross section was completed
\cite{Baernreuther:2012ws, Czakon:2012zr, Czakon:2012pz, Czakon:2013goa}, based on a purely numerical framework for the calculation of the virtual corrections \cite{Czakon:2008zk} and the real radiation subtractions \cite{Czakon:2010td,Czakon:2011ve}. When
these results are supplemented by the resummation of the threshold 
logarithmically-enhanced terms, due to the soft-gluon emission, at the 
next-to-next-to-leading logarithmic (NNLL) accuracy, the perturbative
uncertainty at  LHC center of mass energies is as low as $3 \%$
\cite{Czakon:2013goa}. This is just a few percent smaller than the current
experimental uncertainty. 


NNLO QCD computations require the evaluation of tree-level diagrams with the
emission of two additional partons in the final state (real corrections), of
one-loop  diagrams with an additional parton in the final state (real-virtual
corrections) \cite{Dittmaier:2007wz,Dittmaier:2008uj,Bevilacqua:2010ve,Bevilacqua:2011aa,Melnikov:2010iu}, and of two-loop diagrams without additional partons in the final
state (virtual corrections). Each one of these three elements involves a large
number of diagrams and, in top-quark pair production, it is further complicated
by the need to account for the finite mass of the top-quark. The task of 
combining these elements and organizing the cancellation of the residual
infrared singularities among the various terms in a way which is efficient for
numerical evaluation is itself very challenging \cite{Kosower:1997zr,
GehrmannDeRidder:2005cm, Daleo:2006xa, Daleo:2009yj, Glover:2010im,
Boughezal:2010mc, GehrmannDeRidder:2012ja, Bernreuther:2011jt, Abelof:2011jv, GehrmannDeRidder:2011aa, 
Abelof:2011ap,  Abelof:2012rv,  Abelof:2012he, Weinzierl:2003fx,
Frixione:2004is, Somogyi:2006da, Somogyi:2008fc, Catani:2007vq,
Anastasiou:2003gr, Binoth:2004jv, Anastasiou:2010pw, Czakon:2010td,
Czakon:2011ve, Bierenbaum:2011gg}. 


In this paper we focus our attention on the analytic calculation of the two-loop
corrections to heavy-quark hadroproduction. (In this context we will use the
terms heavy quark and top quark interchangeably.) Even with purely numerical results available, 
there are strong motivations for an analytic calculation. Firstly, it
facilitates a clearer understanding of the structure of the result. Furthermore,
it offers particularly robust, fast and precise numerical evaluations for the
theoretical predictions of the physical observables. Finally, it also provides
an excellent cross check of the numerical calculations.


The two-loop QCD corrections to top-quark pair production were at first
evaluated in the small mass limit \cite{Czakon:2007ej, Czakon:2007wk}. In those
papers,  terms proportional to positive powers of the heavy-quark mass $m$ were
neglected and $m$ was only kept as a regulator of collinear singularities. The
first diagrams to be calculated analytically by retaining the full dependence on
the top-quark mass were the quark-annihilation channel diagrams involving a
closed light- or heavy-quark loop \cite{Bonciani:2008az}. Subsequently, the
diagrams contributing to the leading color structure were evaluated in both the
quark-annihilation \cite{Bonciani:2009nb} and gluon-fusion
\cite{Bonciani:2010mn} channels. The calculations in \cite{Bonciani:2008az,
Bonciani:2009nb, Bonciani:2010mn} were carried out by identifying a set of
Master Integrals (MIs) by means of the Laporta Algorithm \cite{Laporta1,
Laporta2, Tka, Chetyrkin1}, and by subsequently evaluating the MIs using
the Differential Equations Method \cite{Kotikov:1990kg, Kotikov2, Kotikov3,
Remiddi1,Caffo1,Caffo2,Gehrmann1,Argeri}. In order to carry out these
calculations, a process independent, multi-purpose implementation of the Laporta
Algorithm was written in {\tt C++}, the package {\tt Reduze}.
This software employs functionalities from {\tt GiNaC} \cite{ginac} and {\tt Fermat} \cite{fermat} and it is now
publicly available \cite{Studerus:2009ye, vonManteuffel:2012np}.
The final analytic results were written in terms of
Goncharov multiple polylogarithms or Generalized Harmonic Polylogarithms (\gpls) 
\cite{Goncharov1,Goncharov2,Gehrmann:2000zt,Gehrmann2}, which contain the
Harmonic Polylogarithms (HPLs)
\cite{Goncha1,Broad1,Remiddi2,NUMHPL,Vollinga, Maitre:2005uu, Maitre:2007kp, Buehler:2011ev}
as a subset. Expansions near the production threshold and in the small mass
(high energy) limit were obtained for all 
of the results in \cite{Bonciani:2008az, Bonciani:2009nb, Bonciani:2010mn}.


In this paper we present the calculation of the gluon fusion channel two-loop
diagrams involving a light (i.e.\ massless) quark loop. The analytic structure of
the MIs involved in this calculation is more complicated than the one we
encountered in the past. The most complicated MIs, needed for the two-loop box
diagrams, were recently evaluated in \cite{vonManteuffel:2012je,
vonManteuffel:2013uoa, Borowka:2013cma} and require a significantly extended set
of \gpls. The analytic results for the light quark two-loop corrections allow
for a clean extraction of the relevant real parts in the physical region, they
are suitable for stable and fast numerical evaluations, and it is possible to
expand them  both in the production threshold limit $s \to 4 m^2$ and in the
small mass limit  $m^2/s \to 0$. 
In order to achieve all of these goals it was necessary to rewrite the results
in terms of different functional bases, which is a non-trivial task. Symbol and
coproduct based techniques for multiple polylogarithms were already proven to be
powerful tools in various $N=4$ Super-Yang-Mills Theory and QCD applications
\cite{Goncharov:2010jf,Brown:2011ik,Dixon:2011pw,Duhr:2011zq,Dixon:2011nj,
Duhr:2012fh,Dixon:2012yy,Drummond:2012bg,
Chavez:2012kn,vonManteuffel:2012je,Gehrmann:2013vga,
Anastasiou:2013srw,Golden:2013xva,vonManteuffel:2013uoa,
Gehrmann:2013cxs,Henn:2013woa,Dixon:2013eka}. Here, we extensively apply the
coproduct-augmented symbol calculus presented in \cite{Duhr:2011zq,Duhr:2012fh}
for our purpose. Ultimately, the corrections evaluated here are written in terms
of logarithms, classical polylogarithms $\Li_n$ ($n=2,3,4$), and genuine
multiple polylogarithms $\Li_{2,2}$. All of these transcendental functions are
real over the entire physical phase space.

The paper is organized as follows: in Section~\ref{sec:notations} we introduce
our notation and conventions. The calculational method 
is described in Section~\ref{sec:method}, which also includes a description of
different functional bases used to represent our results at different stages of
the calculation. The terms which are proportional to the number of light quarks
$N_l$ in the interference between  the two-loop and tree-level amplitudes  can
be arranged in seven gauge-invariant color coefficients. The analytic
expressions for these coefficients are discussed in Section~\ref{sec:anres}.
These results are too lengthy to be included in the text. For this reason we
collect them in an ancillary file which we attach to the arXiv submission of
this work.  The expansion of the seven color coefficients near the production
threshold and in the small mass limit is presented in Section~\ref{sec:exp}.
Section~\ref{sec:concl} contains our conclusions. In Appendix~\ref{sec:numbers},
we supply numerical values for the considered color coefficients in a few phase
space points.


\section{Notations and conventions}
\label{sec:notations}

In this paper we consider the two-loop corrections to the partonic scattering process
\be
g \left(p_1 \right) + g \left(p_2 \right) \longrightarrow t\left(p_3\right) 
                +  \bar{t}\left(p_4\right) \, , 
\ee
where $p_1^2 = p_2^2 = 0$ and $p_3^2 = p_4^2 = -m^2$. It is convenient to introduce the
Mandelstam invariants, which are defined as follows:
\be
s = -\left(p_1 + p_2\right)^2 \, , \qquad t = -\left(p_1 - p_3\right)^2 \, , 
\quad \text{and} \qquad u = - \left(p_1 - p_4\right)^2 \, .
\label{eq:Mandelstam}
\ee
The invariants satisfy the momentum conservation relation $s + t + u = 2 m^2$.
The squared  matrix element (summed over spin and colour)
can be expanded in powers of the strong coupling constant 
$\alpha_s$ according to
\be
\sum \left| {\mathcal M} \left(s,t,m^2,\varepsilon \right)\right|^2 = 
               16 \pi^2 \alpha_s^2 \left[ {\mathcal A}_0 
	           + \frac{\alpha_s}{\pi} {\mathcal A}_1 
		   + \left( \frac{\alpha_s}{\pi}\right)^2 {\mathcal A}_2 
		   + {\mathcal O}\left(\alpha_s^3\right)\right] \, . 
\label{eq:exp}
\ee

In Eq.~(\ref{eq:exp}), the argument $\varepsilon = (4-d)/2$ indicates the dimensional
regulator. We suppressed the arguments $s$, $t$, $m^2$, $\varepsilon$ in the functions 
${\mathcal A}_i$. It must be remarked that after UV renormalization the terms ${\mathcal
A}_i$  ($i \ge 1$) still include IR divergences, which are regulated by $\varepsilon$.
These divergences cancel only after the virtual corrections are added to the real
emission ones. The term ${\mathcal A}_0$ in Eq.~(\ref{eq:exp}) arises from the
interference of tree-level diagrams; its explicit expression is well known (see for
example \cite{Bonciani:2010mn}). The term ${\mathcal A}_1$ indicates the interference 
of one-loop and tree-level diagrams \cite{Nason:1987xz,Beenakker:1988bq}. The term
${\mathcal A}_2$ can be further split in the sum of two contributions
\be
{\mathcal A}_2 = {\mathcal A}^{(2 \times 0)}_2 +  {\mathcal A}^{(1 \times 1)}_2 \, . 
\label{eq:split}
\ee
${\mathcal A}^{(1 \times 1)}_2$ arises from the interference of one loop diagrams and
was evaluated in \cite{Korner:2008bn,Kniehl:2008fd,Anastasiou:2008vd}.  The color structure of the
interference of two-loop and tree-level diagrams, ${\mathcal A}^{(2 \times 0)}_2$, is
the following
\begin{align}
{\mathcal A}^{(2 \times 0)}_2 = \left( N_c^2-1\right) \Biggl\{ & N_c^3 A  + N_c B + \frac{1}{N_c} C + \frac{1}{N_c^3} D + N_c^2 N_l E_l + N_c^2 N_h E_h + N_l F_l + N_h F_h \nonumber \\
& + \frac{N_l}{N_c^2} G_l + \frac{N_h}{N_c^2} G_h  + N_c N_l^2 H_l + N_c N_h^2 H_h +N_c N_l N_h H_{lh} + \frac{N_l^2}{N_c} I_l \nonumber \\
& + \frac{N_h^2}{N_c} I_h +  \frac{N_l N_h}{N_c} I_{lh} \Biggr\} \, , \label{eq:cocoef}
\end{align}
where $N_c$ indicates the number of colors, $N_l$ the number of light (i.e.\ massless)
flavor quarks and $N_h$ the number  of quarks of mass $m$. In the case of top-quark pair
production one should set $N_h =1$. The 16 gauge-invariant color coefficients $A$, $B$,
$\ldots, I_{lh}$ are functions of $s$, $t$, $m^2$ and $\varepsilon$. To date, only the
leading color coefficient, $A$, was calculated analytically \cite{Bonciani:2010mn}. In
the present work we evaluate the seven color coefficients proportional to $N_l$: $E_l$,
$F_l$, $G_l$, $H_l$, $H_{lh}$, $I_l$ and $I_{lh}$.


\section{Calculational method}
\label{sec:method}

\subsection{Outline of the calculation}

\begin{figure}
\begin{center}
 \includegraphics[height=0.30\textheight]{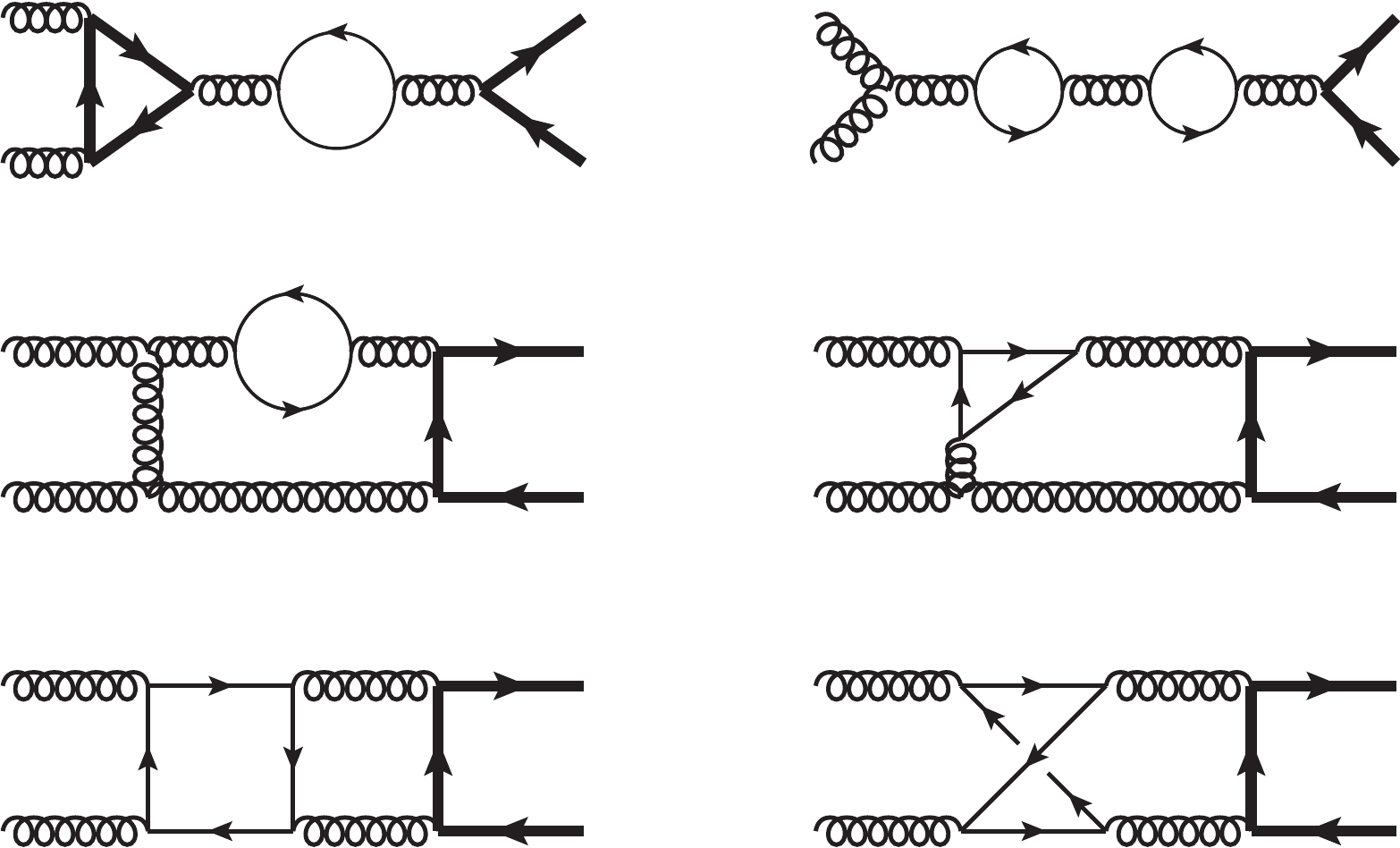}
\end{center}
\caption{\label{fig:diagrams}
Some of the two-loop diagrams needed for the evaluation of the 7 color coefficients discussed in this paper. Thick fermion lines represent heavy  quarks, thin fermion lines represent massless quarks. The diagrams in the first line contribute to color coefficients proportional to $N_l N_h$ and $N_l^2$, respectively. The box diagrams in the last two lines contribute to the color coefficients proportional to $N_l$.
}
\end{figure}

In this section we briefly summarize the way in which the calculation was organized and
carried out. The two-loop Feynman diagrams which contribute to the $gg \to t \bar{t}$
process were generated with {\tt Qgraf} \cite{qgraf}. The total number of diagrams in
this channel is $789$. Diagrams involving a closed light-quark loop are counted only
once and are multiplied by the number of light quarks. $126$ diagrams are proportional
to $N_l$, $6$ are proportional to $N_l N_h$, and $3$ are proportional to $N_l^2$. Some
of the two-loop diagrams which we need to evaluate are shown in
Fig.~\ref{fig:diagrams}. Since in our calculation we set the sum over the gluon
polarization vectors equal to the metric tensor, we need to add the contribution of the
diagrams with incoming Faddeev-Popov ghosts. This amounts to consider $31$ additional
diagrams proportional to $N_l$, $2$ diagrams proportional to the product $N_l N_h$, and
$1$ diagram proportional to $N_l^2$. The {\tt Qgraf} output becomes the input for {\tt
Reduze~2} \cite{vonManteuffel:2012np}. The program starts by interfering the two-loop
diagrams with the tree-level amplitude, and by subsequently calculating traces over
color, spinor and Lorentz indices. By carrying out shifts on the integration momenta,
{\tt Reduze~2} assigns to each diagram a sector of an appropriate integral family
(auxiliary topology).
The list of the nine propagator integral families employed in this calculation is
collected in the ancillary file \verb+integralfamilies.yaml+, included in the arXiv
submission of this paper.
Finally, {\tt Reduze~2} generates the system of Integration by Parts Identities (IBPs)
and solves it, identifying the relevant set of MIs. The program handles multiple
integral families at the same time and deals with crossings of external legs. It
identifies shift relations between different sectors (sector relations) or integrals of
the same sector (sector symmetries). Occasionally, sector symmetries provide
information not included in  the IBP identities. Some of the integrals  needed in the calculation presented here
 were already available in the literature 
\cite{vanNeerven:1985xr,Argeri:2002wz,Bonciani:2003te,Bonciani2,Fleischer1,Aglietti2,
DK,Aglietti:2004tq,CGR,heavytolight1,heavytolight2}. However, 11 integrals with two
massive and two massless external legs had to be calculated specifically for this
project. Their evaluation has many interesting aspects and a considerable amount of
work was necessary in order to bring the final analytic result to a manageable form.
The calculation of these 11 integrals is described in detail in
\cite{vonManteuffel:2013uoa}. In order to obtain the 7 color coefficients we are
interested in, up to terms of ${\mathcal O} (\varepsilon^0)$ included, the MIs need to
be evaluated up to the order in $\varepsilon$ where \gpls\  of weight 4 appear.
 
Once the bare expression of the 7 color coefficients is available, one needs to carry
out the UV renormalization. The renormalization procedure is standard. However, we
provide some further details in Section~\ref{sec:ren}.


\subsection{Renormalization \label{sec:ren}} 

As in our previous papers  on two-loop corrections to top-quark pair production
\cite{Bonciani:2008az, Bonciani:2009nb,Bonciani:2010mn}, we employ a mixed
renormalization scheme in which the wave functions and the heavy-quark mass are
renormalized on shell, while the strong coupling constant is renormalized in the
$\overline{\text{{\small MS}}}$ scheme. The explicit expressions of the one- and
two-loop renormalized amplitudes in terms of bare functions  and counterterms
are explicitly provided in Eq.~(4.6) of reference \cite{Bonciani:2010mn}. As
explained in that work, the bare amplitudes as well as the relevant
renormalization constants are expanded in powers of $\alpha_s/\pi$. The one-loop
renormalization constants can  be found in Eqs.~(4.7-4.10) in
\cite{Bonciani:2010mn}.  Here, we just add the explicit expression of the
two-loop renormalization constants needed to  renormalize the $N_l$, $N_l N_h$
and $N_l^2$ part of the two-loop matrix elements. By employing the notation in
\cite{Bonciani:2010mn} one finds
\begin{align}
\hspace{-2mm} 
\delta Z_{\text{{\tiny WF}}, t}^{(2)} &=C^2\left(\varepsilon\right) \left( \frac{\mu^2}{m^2}\right)^{2 \varepsilon} \! \! 
C_F N_l T_R\left[
\frac{1}{8
   \varepsilon^2}+\frac{9}{16
   \varepsilon}+\frac{\zeta_2}{2
   }+\frac{59}{32}
\right] , \\
\hspace{-2mm} 
\delta Z^{(2)}_{\text{{\tiny WF}}g} & = 0 \, , \\
\hspace{-2mm} 
\delta Z_{\alpha_s}^{(2)} &= C^2\left(\varepsilon\right) \left( \frac{e^{-\gamma \varepsilon}}{\Gamma(1+\varepsilon)}\right)^2 \! \! \frac{N_l T_R}{4 \varepsilon} \left[ \left( \frac{8}{9} N_h T_R + \frac{4}{9} N_l T_R - \frac{22}{9} C_A \right) \! \frac{1}{\varepsilon}
+ \frac{5}{6} C_A + \frac{C_F}{2}  \right] ,\\
\hspace{-2mm} 
\delta Z^{(2)}_m &=C^2\left(\varepsilon\right) \left( \frac{\mu^2}{m^2}\right)^{2 \varepsilon} \! \! \! N_l T_R \left[\frac{1}{8
   \varepsilon^2}+\frac{7}{16
   \varepsilon}+\frac{\zeta_2}{2
   }+\frac{45}{32} + \left( \frac{279}{64} +\frac{7}{4} \zeta_2 + \zeta_3 \right) \varepsilon \right] . 
\end{align}
As usual, in the QCD case with $\text{SU}(N_c=3)$ we have $C_A = 3$, $C_F = 4/3$ and $T_R = 1/2$. Finally, $\gamma \approx 0.577216$ is the Euler-Mascheroni constant and $C(\varepsilon) = (4 \pi)^\varepsilon \Gamma(1+\varepsilon)$ is a factor which reabsorbs the logarithms of $4 \pi$ and $\gamma$-dependent terms in the $\varepsilon$ expansion.


\subsection{Multiple polylogarithms \label{sec:vodoo}}

Our results are obtained by inserting the appropriate MIs in the IBP reduction of each
diagram. Consequently, the results share the analytical properties of the MIs,
which have different branch cuts.
The correct way of crossing the cuts is dictated  by causality, which is enforced by
the $i \delta$ ($\delta \to 0^+$) term in the Feynman propagators.
The various MIs encountered in the present calculation can have thresholds located at
$s=0$, $s =4 m^2$, $t = m^2$ and $u = m^2$ (see the discussion in  \cite{vonManteuffel:2013uoa}).
In the physical region one finds that $s \ge 4 m^2$ and
that both $t$ and $u$ are negative. It is then sufficient to associate an infinitesimal
positive imaginary part to $s$ in order to have  well defined results. 

The color coefficients $E_l$, $F_l$, $G_l$, $H_l$, $H_{lh}$, $I_l$, and $I_{lh}$ depend 
on three independent variables: the top-quark mass $m$ and two of the three 
dimensionless variables $x$, $y$ and $z$, where
\be
x = - \frac{1-\sqrt{1-\frac{4 m^2}{s}}}{1+\sqrt{1-\frac{4 m^2}{s}}} \, , \qquad y = -\frac{t}{m^2} \, ,\qquad z = -\frac{u}{m^2} \, .
\ee
The momentum conservation relation among Mandelstam invariants can be rewritten in
terms of $x$, $y$ and $z$ as follows:
\be
y +z + \frac{1+x^2}{x} = 0\, . 
\label{eq:relxyz}
\ee
In the physical phase-space region, the following inequalities  hold:
\be
m^2 > 0 \, , \quad  -1 \le x < 0 \, , \quad -x \le y \le -\frac{1}{x} \, , 
\quad -x \le z \le -\frac{1}{x} \, , \quad yz \ge 1 \, , \quad
y +z \ge 2 .
\ee

The color coefficients we are interested in can be written in terms of \gpls\  of
arguments $x$, $y$ and $z$ up to weight four \cite{vonManteuffel:2013uoa}. The \gpls\  
are defined through iterated integrations by the relations
\begin{align}
G\left(w_1, w_2, \cdots, w_n; q \right) &\equiv \int_0^q dt \, 
\frac{1}{t -w_1} G\left(w_2, \cdots, w_n; t \right) \, ,
\nonumber \\
G\bigl( \underbrace{0,\cdots, 0}_n; q\bigr) &\equiv \frac{1}{n!} \ln^n q \,.
\end{align}
The variable $q$ can be  $x$, $y$ or $z$ and the weights $w_i$ are 
simple rational functions of $x$, $y$, $z$ or complex constants ($w_i \in \mathbb{C}$). 

For univariate multiple polylogarithms with argument $q=x$, it was found useful to generalize
the definition of \gpls\  by allowing polynomial denominators, as follows
\be
G\left( [f(o)], w_2, \cdots, w_n; q\right) = \int_0^q dt \,  \frac{f'(t)}{f(t)} G\left( w_2, \cdots, w_n; t\right) \, ,
\label{eq:Gf}
\ee 
where $f(o)$ is an irreducible rational polynomial of arbitrary degree
and $o$ is a dummy variable.
Here, the weights are restricted to be independent of external variables.
In the following, we will refer to the weights of the form $[f(o)]$ as {\em
generalized weights}~\cite{vonManteuffel:2013vja,vonManteuffel:2013uoa}.
Since $f(o)$ is a polynomial of degree $n$, one can always choose
to normalize the coefficients in such a way that the cofactor of $o^n$ is equal to $1$.
The polynomial $f(o)$ can be written in terms of its $n$ complex roots $r_i$ as
\be
f(t) = (t-r_1) (t-r_2) \cdots (t-r_n) \, ;
\ee
since
\be
\frac{f'(t)}{f(t)} = \frac{1}{t -r_1} + \frac{1}{t -r_2} + \cdots +  \frac{1}{t -r_n}\, ,
\ee
a \gpl\ with generalized weights is related to \gpls\ with constant complex weights $r_i$
\be
G\left(\cdots, [f(o)], \cdots; q \right) =
G\left(\cdots \!, r_1, \cdots \!; q \right) \! 
+ \! G\left(\cdots \!, r_2, \cdots \!; q \right) \!
+ \! \cdots \! + \! G\left(\cdots \!, r_n, \cdots \!; q \right)\,.
\label{eq:relGf}
\ee
In our case, the introduction of the \gpls\  in Eq.~(\ref{eq:Gf})
is indeed enough to handle all non-linear denominators in the integrating factors
and no complex weights are needed.
In particular, we employ a single \gpl\  with weight $[f(o)] = [1-o+o^2]$ instead of the
corresponding pair of \gpls\  with complex weights
\be
r_{1,2} = -\frac{1}{2} \left(1 \pm i \sqrt{3} \right) \, ,
\ee
using Eq.~(\ref{eq:relGf}).
Similarly, we encounter the generalized weight $[f(o)] = [1+o^2]$, which
is related to the pair of complex weights $r_{1,2} = \pm i$.
Besides leading to more compact results, the generalized weights prevent the appearance of
spurious imaginary parts present in each of the \gpls\  with complex weights.
The \gpls\  of the kind in Eq.~(\ref{eq:Gf}) can be evaluated numerically  by means of the package described
in \cite{Vollinga}, again by employing the relation in Eq.~(\ref{eq:relGf}).
A very important feature of the generalized weights resides in the fact that the corresponding integrating factors satisfy the relation
\begin{equation}
  \mathrm{d}t\,\frac{f'(t)}{f(t)} = \mathrm{d}\ln(f(t))\,,
\end{equation}
which allows for the construction of  symbols and coproducts
without reference to roots of irreducible polynomials
(see \cite{vonManteuffel:2013vja} for details and
\cite{Ablinger:2011te,Bogner:2012dn} for related extensions
of multiple polylogarithms).

A large fraction of the work presented in this paper was devoted to simplify the color coefficients in terms of \gpls,
cast them into a representation suitable for numerical evaluation
and expand them in different kinematical limits.
Initially, the MIs are expressed in terms of different pairs of variables chosen in the
set $\{x,y,z\}$.
For a planar integral there is usually a ``natural'' choice for this pair (corresponding
to the cut structure of the MI), which renders the expressions particularly compact.
The situation is considerably more involved for non-planar integrals, where
cuts in all three channels $s$, $t$ and $u$ are present simultaneously.
A naive substitution of the MIs in the expression of the cross section produces large analytic results which are difficult to manipulate.
The difficulties are of several kinds. First of all, the simultaneous  presence of $x$, $y$ and $z$ 
gives rise to redundancies in the representation of the color coefficients (i.e.\ the latter can contain hidden 
zeros), because of Eq.~(\ref{eq:relxyz}).
This proliferation of terms makes the numerical evaluation of the color coefficients more time consuming.
Moreover, it is a non-trivial task  to extract the real and imaginary parts of the
interference between two-loop and tree-level diagrams and to write it in terms of real-valued \gpls.
In fact, we encounter \gpls\  which are real in some phase-space regions and 
complex in others.
Finally, \gpls\  of arguments $x$, $y$ and $z$ are not straightforwardly  expanded
near the production threshold or in the asymptotic small-mass (or high-energy) limit.
Our solution to these problems relies on the  methods outlined below.

At first, we choose to rewrite the color coefficients by eliminating $z$ in terms of $x$ and $y$. Furthermore,  for 
$y$-dependent \gpls\  we choose $y$ as the \gpl\ argument (resulting in $x$-dependent or constant weights), while
for $y$-independent \gpls\  we choose $x$ as argument (resulting in generalized or constant weights).
Involved substitutions are necessary in order to recast the \gpls\ depending on $\{z,x\}$ or $\{y,z\}$ in terms \gpls\ depending on 
$\{y,x\}$.
These transformations as well as others needed for the expansions in the kinematic limits discussed in Section~\ref{sec:exp} are generated by means of {\tt GPLChangeArg} \cite{AndreasGPL},
a new automated argument-change algorithm written in {\tt Mathematica}. The algorithm does not involve the use
of the symbol or the coproduct and matches constants using high precision numerical evaluations of \gpls~\cite{Vollinga}.

It turns out that the color coefficients are faster to evaluate and numerically more stable when we
insist on the use of specific types of multiple polylogarithms rather than on specific ways in which the kinematic invariants
$x$ and $y$ enter in the argument and weights of the transcendental functions.
For this reason, we rewrite our results in a  second normal form. In this case, 
instead of \gpls\  of arguments $x$ and $y$, we choose to employ the functions $\ln$, $\Li_n$ ($n=2,3,4$) and $\Li_{2,2}$,
where the arguments are (complicated) rational functions of $x$ and $y$.
In particular, we choose the arguments such that the functions are real valued everywhere
in the physical region of phase space.
The $\Li_{m_1,\cdots,m_k}$ functions ($m_i\in\mathbbm{N}$) can be
represented in terms of nested sums
\be
\Li_{m_1, \cdots, m_k}(x_1, \cdots, x_k) = \sum_{i_k=1}^\infty   \frac{x_k^{i_k}}{i_k^{m_k}} \sum_{i_{k-1} = 1}^{i_k-1} \frac{x_{k-1}^{i_{k-1}}}{i_{k-1}^{m_{k-1}}} \cdots \sum_{i_1 = 1}^{i_{2}-1}\frac{x_1^{i_1}}{i_1^{m_1}}  \, ,
\ee
for $|x_1 \cdots x_j| \le 1$ for all $j \in \{1,\cdots, k\}$ and $(m_1,x_1) \neq (1,1)$.
In general, $\Li$ functions and \gpls\  are related by
\be
\Li_{m_1, \cdots, m_k}(x_1,\ldots,x_k) = (-1)^k G\bigg(\underbrace{0,\ldots,0,\frac{1}{x_k}}_{m_k},\ldots,\underbrace{0,\ldots,0,\frac{1}{x_1\cdots x_k}}_{m_1}\bigg)
\ee
and thus describe the same set of functions.
In particular, they can be evaluated numerically with the same tools employed for \gpls\ \cite{Vollinga}.
Explicitely, we have
\be
\Li_n (x) = -G\big(\underbrace{0,\cdots,0,1}_{n};x\big)   \, , \qquad 
\Li_{2,2}(x_1,x_2) = G\left(0, \frac{1}{x_1}, 0, \frac{1}{x_1 x_2}; 1\right) \,.
\ee
for the functions considered here.
The translation to the new functional basis is carried out by employing another automated
algorithm implemented in {\tt Mathematica}: {\tt GPLReduce} \cite{AndreasGPL}.
The latter is an improved coproduct based normal form algorithm, and it
extends the procedure outlined in \cite{Duhr:2011zq} to the case of \gpls\ depending on
generalized weights.
The algorithm determines algebraic constants
from numerical samples of \gpls\ obtained with~\cite{Vollinga}.
It is indeed interesting to
observe that algorithms based on coproduct apply also to the case of the generalized weights
introduced in Eq.~(\ref{eq:Gf}).


\section{Analytic results}
\label{sec:anres}

\begin{figure}[!ptb]
\centerline{\begin{tabular}{rr}
 \includegraphics[height=0.222\textheight]{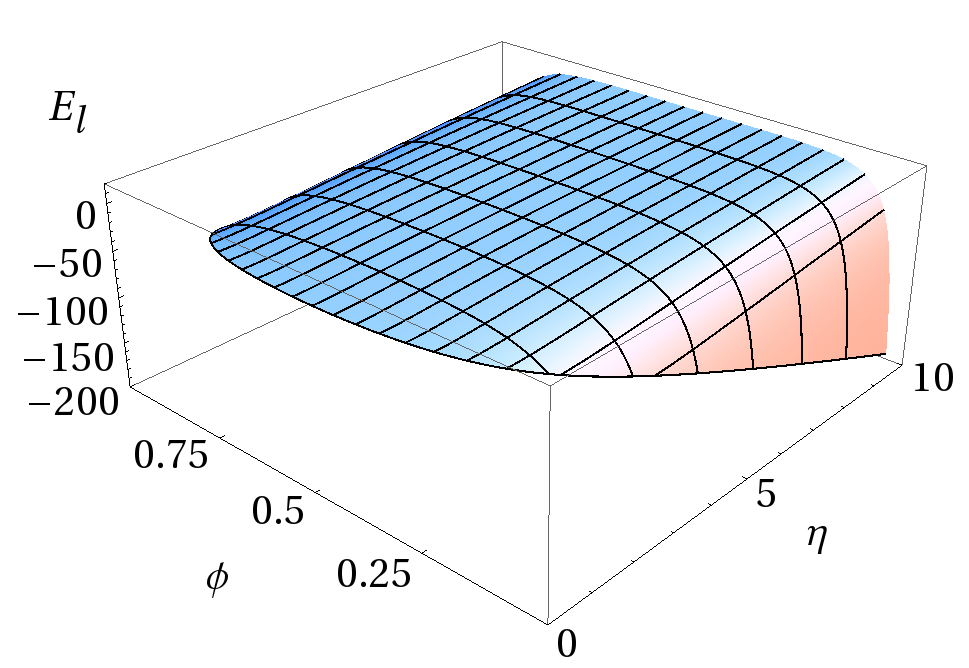}&
 \includegraphics[height=0.222\textheight]{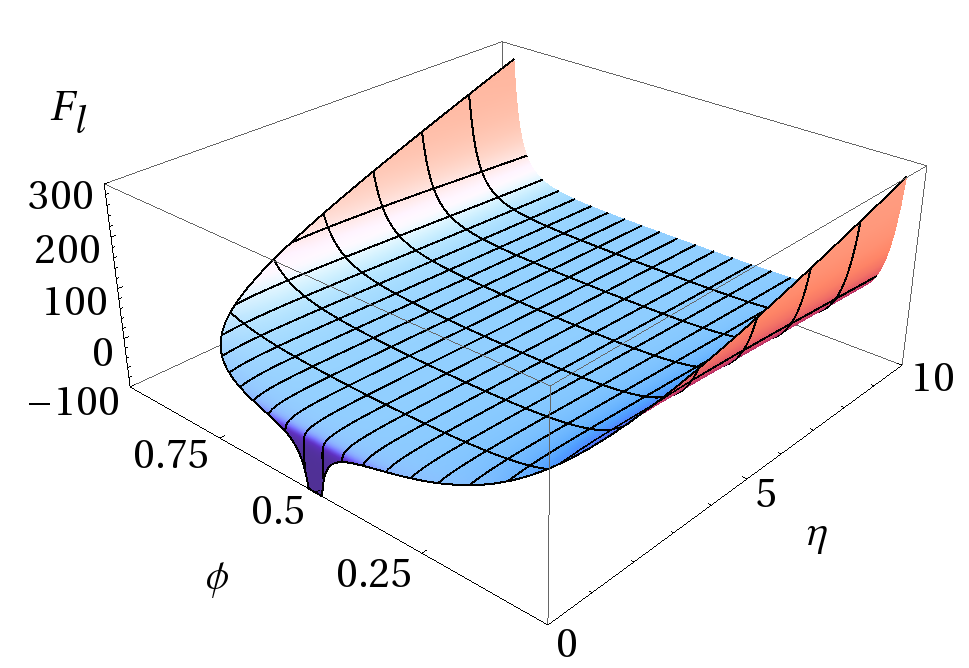}\\[-1ex]
 \includegraphics[height=0.222\textheight]{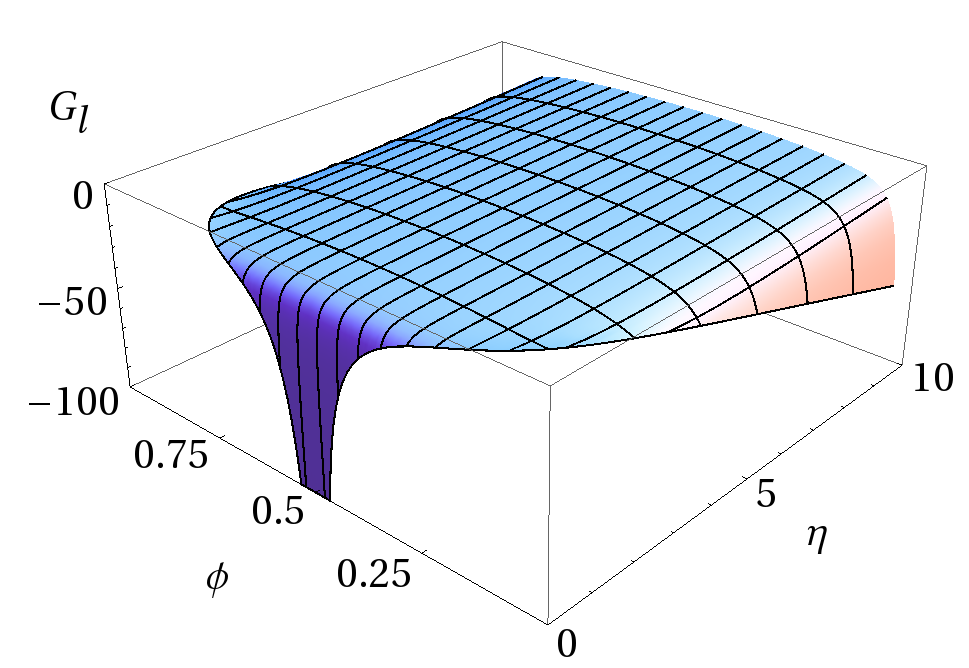}&
 \includegraphics[height=0.222\textheight]{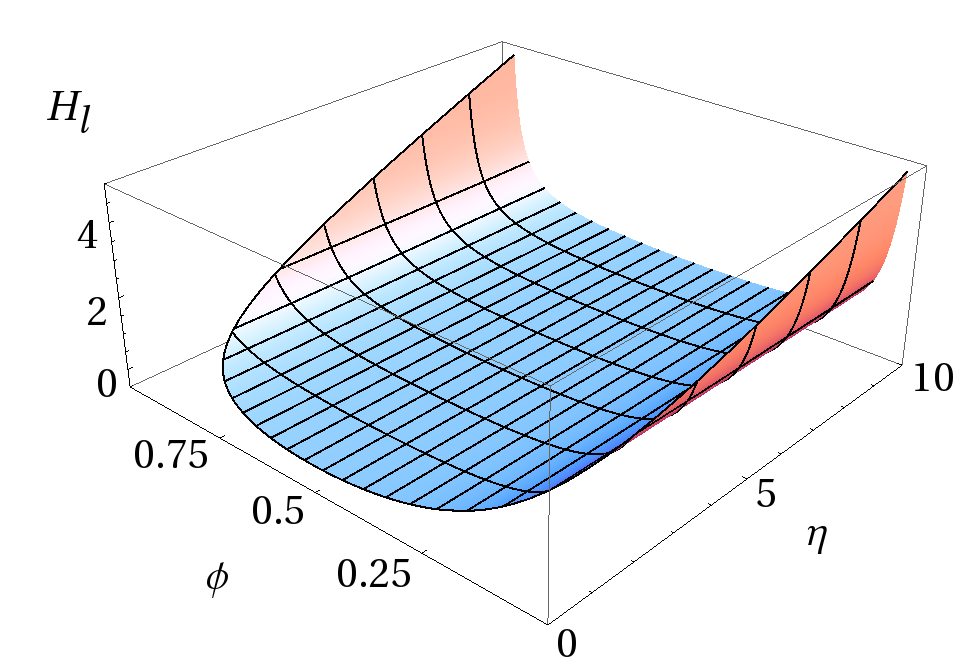}\\[-1ex]
 \includegraphics[height=0.222\textheight]{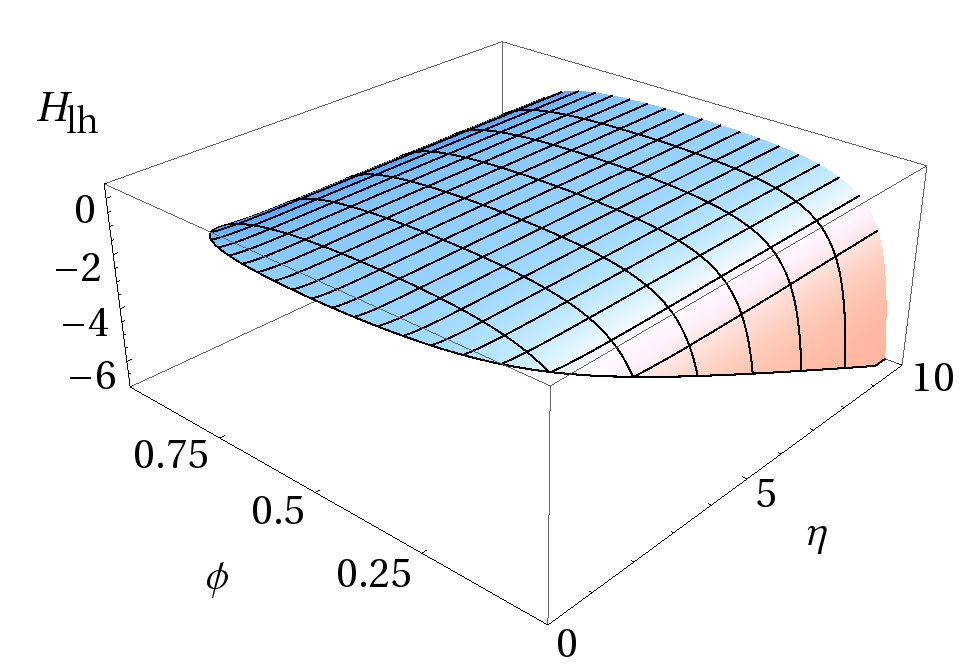}&
 \includegraphics[height=0.222\textheight]{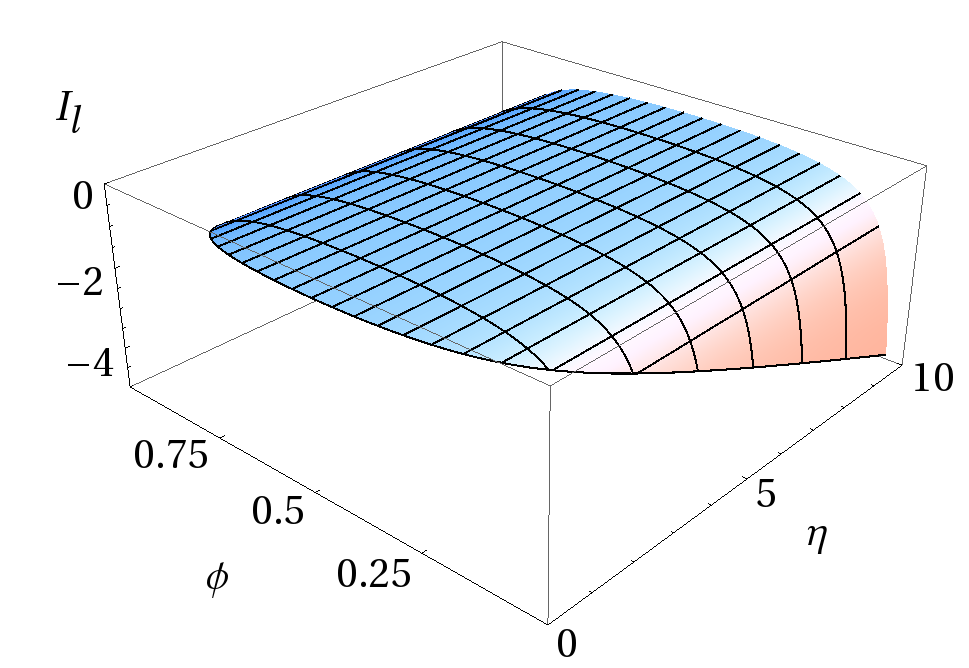}\\[-1ex]
 \includegraphics[height=0.222\textheight]{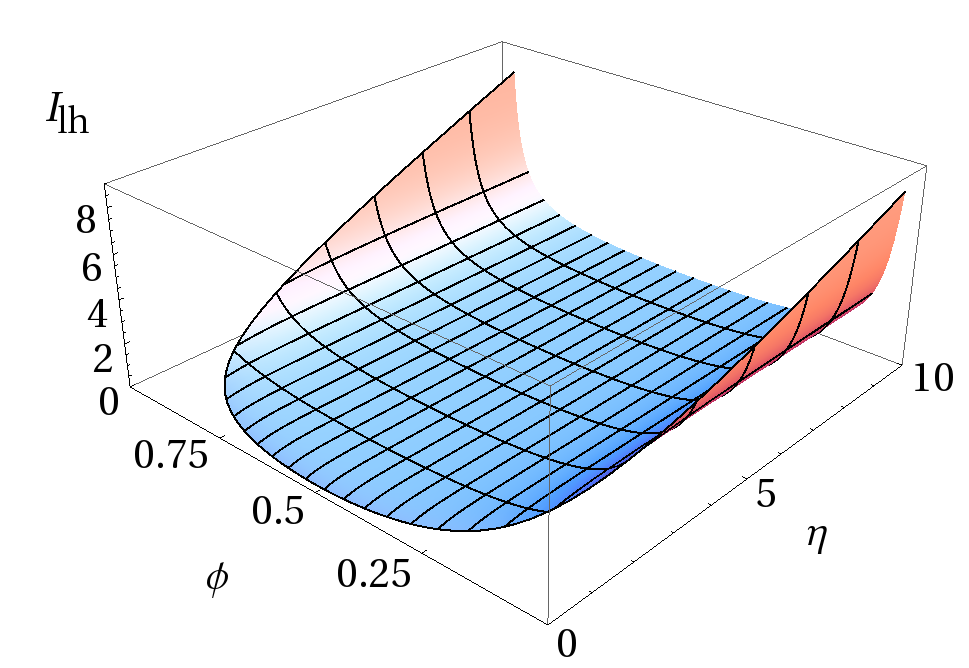}&
\end{tabular}}
\caption{\label{fig:etaphi3d}
Finite parts of the color coefficients as functions of $\eta=s/(4m^2)-1$ and 
$\phi=-(t-m^2)/s$. The finite parts plotted here are normalized in the standard 
$\overline{\mbox{{\small{MS}}}}$ way (see Appendix~\ref{sec:numbers}). The renormalization 
scale $\mu$ is set equal to the top quark mass.}
\end{figure}

In this section we present our results. By using the techniques described in
Section~\ref{sec:vodoo}, we can write  closed  analytic expressions for the UV
renormalized color coefficients $E_l$, $F_l$, $G_l$, $H_l$, $H_{lh}$, $I_l$, and $I_{lh}$ in Eq.~(\ref{eq:cocoef}). 
These expressions are valid for arbitrary values of the top-quark mass and of the
Mandelstam invariants. Furthermore, as mentioned in the previous section, they are at first derived by 
writing all the \gpls\ 
in terms of the dimensionless variables $x$ and $y$ (we remind the reader
that $x<0$ in the physical region). 
In the case of \gpls\  of two variables, we always choose $y$ as the functional argument.
Correspondingly, the weights of the \gpls\  of argument $y$ are part of the set
\be
\label{eq:weightsy}
\left\{ -1, 0, -\frac{1}{x}, -x, -\frac{(1 + x^2)}{x}, -\frac{(1 - x + x^2)}{x}\right\} 
\, .
\ee
The \gpls\ of argument $x$ have weights
\be
\label{eq:weightsx}
\left\{ -1, 0, 1, [1 + o^2], [1 - o + o^2]\right\} \, .
\ee
The maximum transcendentality of the \gpls\ present in our results is 4. The
total number of \gpls\ is 289 if we expand  all  \gpls\ products; by exploiting
the shuffle algebra we reduced this number to 221 independent  \gpls. It is
interesting to observe that in the color coefficients $E_l$, $F_l$, and $G_l$
the transcendentality of the \gpls\ present in the finite part is 4, while  the
single pole part involves only \gpls\ of transcendentality 2. The analytic
expressions are lengthy and can be found in the ancillary file
\verb+ggtt-lightnf-argyx.m+ attached to the arXiv submission of this article.

As discussed in Section~\ref{sec:vodoo}, the numerical evaluation speed and stability
of the results benefit from writing the analytic color coefficients  in terms of
specific real valued $\Li$ functions, where the arguments are (complicated) rational functions
of $x$ and $y$.
With this choice, the imaginary parts of the tree-level times two-loop
interferences appear explicitly multiplied by a factor $i \pi$ and it is therefore
trivial to extract the real part of that result, which coincides up to a factor of $2$ 
with the color coefficients we are interested in.
The total number of independent multiple polylogarithms we employ for the seven color
coefficients is 225. Only 57 of these are $\Li_{2,2}$, while the others are logarithms 
and classical polylogarithms $\Li_n$ ($n =2,3,4$).
Examples for the $\Li$ functions' arguments are
\begin{displaymath}
\pm x ,\,  \pm x^2 ,\,  -\frac{1}{y},\,  -y,\,  -\frac{y}{x},\,  -x (x+y),\,  \frac{x+y}{y},\,  -\frac{x+z(x,y)}{x+y}, \cdots \, .
\end{displaymath}
Note that, although the total number of independent functions is not reduced by
employing this representation instead of the one based on \gpls\ of simple arguments,
the numerical evaluation of the color coefficients is a factor of $\sim$ 10-15 faster
for typical phase space points.
We conclude this is due to the simpler functional structure of the
remaining multiple logarithms.
When written in terms of $\Li$ functions, the seven color coefficients can be evaluated numerically in a generic phase-space point in a time of $\mathcal{O}(1s)$.
For these numerical evaluations, we employed one 3.4~GHz core, double precision accuracy, {\tt Mathematica} for the evaluation
of $\ln$ and $\Li_n$ functions and \cite{Vollinga} for the evaluation of $\Li_{2,2}$ functions.

We emphasize that for our representation of the seven color coefficients in
terms of $\Li$ functions, no new symbol letters were introduced beyond the ones
that follow from Eqs.\ (\ref{eq:weightsy}) and (\ref{eq:weightsx}). We observe
that all of the generalized weights $[f(o)]$ introduced in Eq.~(\ref{eq:Gf})
could be eliminated from the result with the new choice of basis functions.
Specifically, the functions depending on the generalized weights appear in the
color coefficients only in particular combinations with other multivariate
polylogarithms. These combinations can be rewritten in terms of normal \gpls\ of
rational weights.

Also in the case in which one employs $\Li$ functions, the complete analytic
results are lengthy. We provide them via the ancillary file
\verb+ggtt-lightnf.m+ attached to the arXiv submission of this work. Finally,
following what was done in \cite{Bonciani:2009nb, Bonciani:2010mn}, in 
Fig.~\ref{fig:etaphi3d} we plot the finite part of the color coefficients as a
function  of the variables $\eta$ and $\phi$, defined as 
\be
\eta  = \frac{s}{4 m^2} -1, \quad
\phi = -\frac{t-m^2}{s}, \quad \text{where}~
\frac{1}{2} \left( 1- \sqrt{\frac{\eta}{1+\eta}}\right) \le \phi \le \frac{1}{2} \left( 1+ \sqrt{\frac{\eta}{1+\eta}} \right).
\ee
In addition, we give numerical values for three benchmark points in Appendix~\ref{sec:numbers}.

We performed several checks of our results. The IR poles of $E_l$, $F_l$, $G_l$,
$H_l$, $H_{lh}$, $I_l$, and $I_{lh}$ were first derived in
\cite{Ferroglia:2009ii}. We found full analytical agreement between the IR poles
we obtained and the results available in the literature. In order to test the
finite parts of the seven color coefficients, we compared them with the
numerical results presented in Table 4.2 of \cite{BaernreutherTh}, that are
valid for $s/m^2=5$, $t/m^2=-1.25$, and $\mu/m=1$. After accounting for the
different normalization, we found complete agreement on all of the $10$
significant digits provided in that table.


\section{Expansions \label{sec:exp}}

One of the advantages of working with analytic expressions resides in the fact
that one can readily obtain useful expansions of the results in particularly
interesting phase-space regions. Here, we consider two different expansions:
the expansion near the production threshold $s \sim 4 m^2$, and the expansion in
the small-mass (high-energy) limit $s \gg m^2$.

\subsection{Threshold expansion}

\begin{figure}[th!]
\centerline{\includegraphics[width=\textwidth]{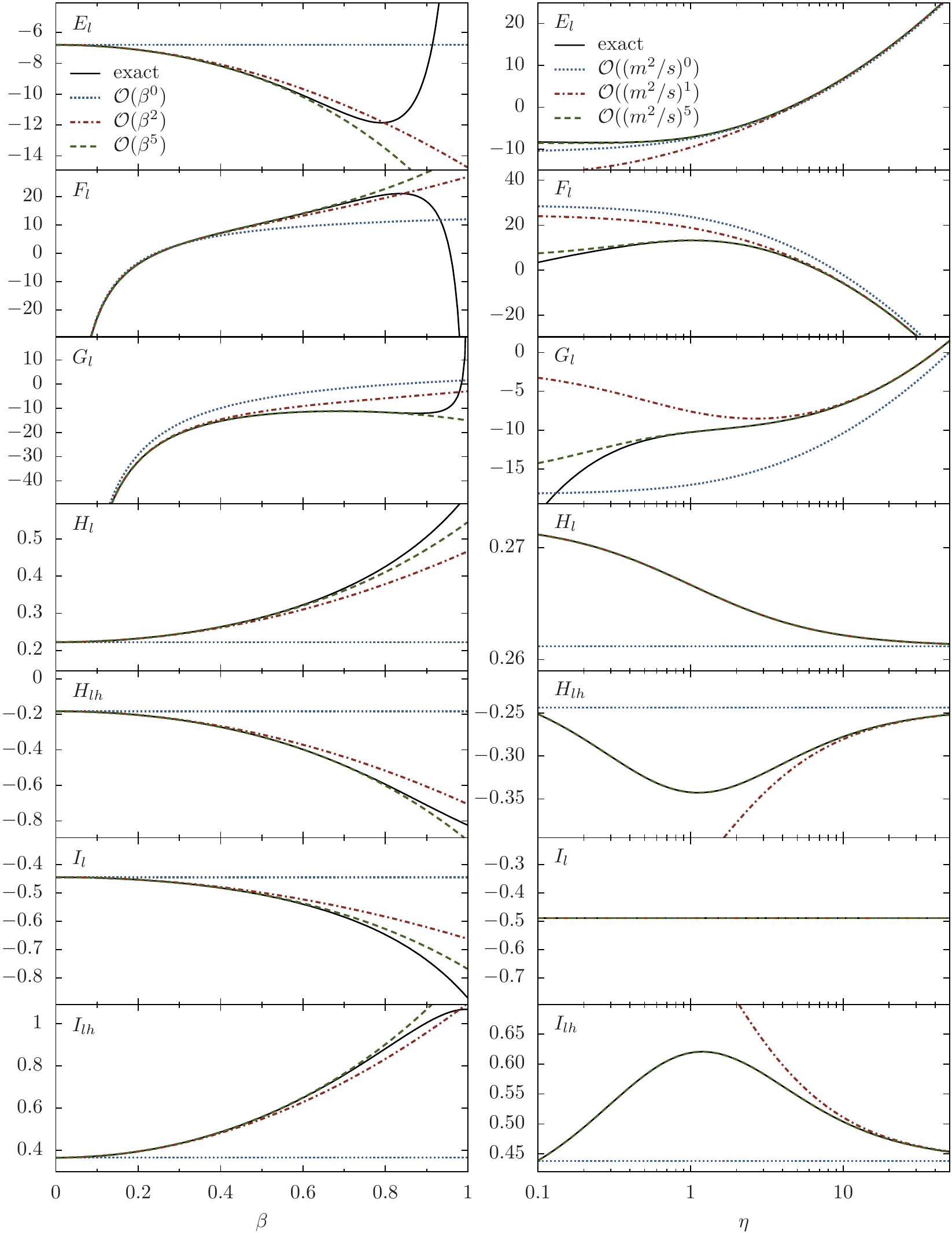}}
\caption{\label{fig:expansions}
Left column: finite parts of the color coefficients as a function of 
$\beta=\sqrt{1-4m^2/s}$ for $\cos\theta=0.7$.
Right column: the same, but in the high energy region as a function of $\eta=s/(4m^2)-1$
for $\phi=-(t-m^2)/s=0.35$.
We set $\mu=m$ and employ $\overline{\text{{\small MS}}}$ normalization
(see Appendix~\ref{sec:numbers}).
}
\end{figure}

The threshold region is identified by taking the limit $\beta \to 0$, where
$\beta$ is the top-quark velocity in the $t \bar{t}$  rest frame,
\be
\beta \equiv \sqrt{1 - \frac{4 m^2}{s}} \, ,
\ee
at fixed scattering angle $\theta$.
The first few orders of the expansion of the color coefficients
take the following remarkably simple form
\bea
E_l & = & - \frac{7}{12 \varepsilon^3}
          - \frac{1}{\varepsilon^2} \left(
                \frac{L_\mu}{2} - \ln2 - \frac{5}{12} \right)
	  +\frac{1}{216 \varepsilon}\bigl(  
	       -36 L_\mu^2+144 L_\mu \ln2+348 L_\mu-144 \ln^2 2 \nn\\ 
& &
      - 696 \ln2 + 252 \zeta_2 + 761 \bigr)
      + \frac{1}{648} \bigl(
           504 L_\mu^2 - 2016 L_\mu \ln2 + 216 L_\mu \zeta_2 - 456 L_\mu \nn\\ 
& &
      + 2016 \ln^2 2 - 432 \zeta_2 \ln 2 + 1884 \ln 2 - 2430 \zeta_2 - 414\zeta_3
      - 3185 \bigr) \nn\\ 
& &
      + \beta^2 \biggl[ 
          - \frac{7}{12\varepsilon^3} (\Xxi+2)
	  + \frac{1}{12\varepsilon^2} \bigl(
	       -6 L_\mu \Xxi-12 L_\mu+12 \ln 2 \Xxi+24 \ln 2+16 \Xxi-23 \bigr) \nn\\ 
& &
      + \frac{1}{216 \varepsilon} \bigl(
         -36 L_\mu^2 \Xxi-72 L_\mu^2+144 L_\mu \Xxi \ln2 +288 L_\mu \ln 2+528 L_\mu \Xxi
	 + 156 L_\mu \nn\\ 
& &
      -144 
      \Xxi\ln^22 -288
      \ln^2 2-1824 
      \Xxi\ln2 +24 \ln2+252
      \Xxi \zeta_2+2513
      \Xxi 
      + 504 \zeta_2\nn\\ 
& &
       -326\bigr) +\frac{1}{648} \bigl(
            720 L_\mu^2 \Xxi+360 L_\mu^2-2880 L_\mu \Xxi \ln2 -1440 L_\mu \ln2 
       + 216 L_\mu \Xxi \zeta_2\nn\\ 
& &
      +228 L_\mu \Xxi+432 L_\mu \zeta_2-696 L_\mu
      + 3456 \Xxi \ln^2 2+864 \ln^2 2
      + 10422 \Xxi \zeta_2 \ln2 \nn\\ 
& &
      +24312 \Xxi \ln2-8316 \zeta_2\ln2-11712 \ln 2 
      - 10476 \Xxi \zeta_2 - 9459 \Xxi \zeta_3 -12200 \Xxi \nn\\ 
& &
      +4158 \zeta_2+5382 \zeta_3 + 2576 \bigr) 
         \biggr]  + {\mathcal O}(\beta^4)\, , \label{eq:Elth} \\
F_l & = & \frac{1}{\beta} \left[ 
            \frac{\zeta_2}{2 \varepsilon} - \frac{7 \zeta_2}{3} \right] 
      + \frac{7}{6 \varepsilon^3} + \frac{1}{\varepsilon^2} \left(
           L_\mu-2 \ln2-\frac{1}{3} \right)
      + \frac{1}{216 \varepsilon}\bigl(
           72 L_\mu^2-288 L_\mu \ln2 \nn\\ 
& &
            -624 L_\mu+288
            \ln^22+1248
            \ln2-450
            \zeta_2-1897\bigr)+\frac{1}{648}
      \bigl(-1008 L_\mu^2+4032
      L_\mu \ln2
\nonumber \\ 
& &   - 432
      L_\mu \zeta_2-96
      L_\mu-4032
      \ln^22+864 
      \zeta_2 \ln2-4128
      \ln2+5103 \zeta_2+693
      \zeta_3 +7339\bigr)
\nonumber \\ 
& &
      +\beta \left[ \frac{1}{\varepsilon}\biggl(\frac{3
               \zeta_2}{2}-\frac{\Xxi
               \zeta_2}{2}\biggr)-\frac{\Xxi
         \zeta_2}{6}-4 \zeta_2 \right]+ {\mathcal O}(\beta^2)\, , \label{eq:Flth} \\
G_l & = & \frac{1}{\beta} \left[ 
              \frac{\zeta_2}{\varepsilon}-\frac{14 \zeta_2}{3} \right] 
      + \frac{1}{\varepsilon}\left(\frac{\zeta_2}{2}-\frac{17}{12} \right)
      + \frac{1}{36} \bigr( 
            216 \zeta_2\ln2+480 \ln2-273 \zeta_2
      - 6 \zeta_3+211 \bigl) 
\nn\\ 
& &
      + \beta \left[
           \frac{3 \zeta_2}{\varepsilon}-4 \Xxi \zeta_2-8 \zeta_2 \right] 
      + {\mathcal O}(\beta^2)\, , \label{eq:Glth} \\
H_l & = & \frac{1}{9 \varepsilon^2}
      - \frac{1}{3  \varepsilon}
      + \frac{1}{9} \left( 2-\zeta_2 \right) 
      + \beta^2 \left[
            \frac{1}{9\varepsilon^2} \bigl( \Xxi + 2 \bigr)
      - \frac{13}{18 \varepsilon} \Xxi
      - \frac{1}{18} \bigl( 2 \Xxi \zeta_2 
      - 9 \Xxi+4 \zeta_2 \bigr) \right] \nn\\
& & 
      + {\mathcal O}(\beta^4)\, , \\
H_{lh} & = & - \frac{2 L_\mu}{9\varepsilon} 
      + \frac{1}{9}\left(
            - L_\mu^2+6 L_\mu-\zeta_2\right) 
      + \beta^2 \biggl[
              \frac{1}{36\varepsilon} \left(
	          14 \Xxi-8 L_\mu \Xxi-16 L_\mu-9 \Xxi \zeta_2\right) \nn\\
& & 
     - \frac{1}{108} \bigl( 
         12 L_\mu^2 \Xxi
       + 24 L_\mu^2
       - 156 L_\mu \Xxi
       + 54 \Xxi \zeta_2 \ln2
       - 84 \Xxi \ln2
       - 33 \Xxi \zeta_2
       + 63 \Xxi \zeta_3
\nonumber \\ 
& &
       + 10 \Xxi 
       + 24 \zeta_2 \bigr) \biggr]
     + {\mathcal O}(\beta^4)\, , \\
I_{l} & = &  -\frac{2}{9 \varepsilon^2}
      + \frac{2}{3 \varepsilon} + \frac{2}{9} \left( \zeta_2 - 2 \right) 
      + \beta^2 \left[
            - \frac{4}{9 \varepsilon^2}+\frac{8}{9 \varepsilon} \Xxi
	    - \frac{4}{9} \left( \Xxi-\zeta_2 \right) \right] 
      + {\mathcal O}(\beta^4)\, , \\
I_{lh} & = &  \frac{4}{9 \varepsilon} L_\mu 
      + \frac{2}{9} \left( \zeta_2 -  6 L_\mu  + L_\mu^2 \right) 
      + \beta^2 \left[ 
              \frac{8 L_\mu}{9 \varepsilon}+\frac{4}{9} \left(\zeta_2 -4 L_\mu \Xxi 
	    + L_\mu^2 \right) \right] 
      + {\mathcal O}(\beta^4)\, , \label{eq:threshold}
\eea
where $L_\mu = \ln(\mu^2/m^2)$  and $c_\theta\equiv\cos\theta$.
The threshold expansion up to and including terms of  $\mathcal{O}(\beta^5)$ can
be found in the ancillary file \verb+ggtt-lightnf-smallbeta.m+, which is
included in the arXiv submission of this work. A comparison between exact
results and threshold expansions at different orders in $\beta$ is shown in
Fig.~\ref{fig:expansions}. In the figure we set $\cos\theta = 0.7$ and $\mu =
m$.

In order to obtain the expansions in Eqs.~(\ref{eq:Elth}-\ref{eq:threshold}), we
start from  the color coefficients written in terms of \gpls\ depending on $x$ and $y$. Subsequently, we rewrite each of the
\gpls\ of arguments $\{y,x\}$ in terms of \gpls\ of arguments $\{\beta,\xi\}$, where $\xi= (1-c_\theta)/2$.
We obtain \gpls\ of argument $\beta$ and weights included in the set
\begin{displaymath}
\left\{-1, 0, 1, \frac{1}{1- 2 \xi},-\frac{1}{1- 2 \xi} , [1 + o^2], [3 + o^2], 
 [1 + 2 o (1 - 2 \xi) + o^2], [1 - 2 o (1 - 2 \xi) + o^2]\right\} \, ,
\end{displaymath}
as well as HPLs of argument $\xi$ and weights $\{0, 1\}$. (The $\xi$ dependence
of the generalized weights is beyond the scope for which we introduced this
extension in Section \ref{sec:vodoo}; this is not an issue here since we do not
consider variations of these \gpls\  with respect to\ $\xi$ nor do we employ
symbols calculus for them.) The translation of the \gpls\ from the $\{y,x\}$ to
the $\{\beta,\xi\}$ basis involves a relatively large number of irrational
constants related to logarithms and $\Li_n$ of fixed argument, as well as
constants which are obtained from \gpls\ of weights $\{-2, -1, 0, 1, 2, [1 +
o^2], [1 + o + o^2]\}$ and argument $1$. These \gpls\ of argument $1$ can be
rewritten in terms of the irrational constants. In some cases the relations can
be easily found. For example:
\begin{align}
G([1+o^2];1)&=\ln2\, ,\qquad G([1+o+o^2];1)=\ln3 \, , \nonumber \\
G(0,[1+o^2];1) &=\frac{\pi^2}{24} \, , \qquad G(0,[1+o+o^2];1)=\frac{\pi^2}{9} \, . \label{eq:Gat1}
\end{align}
However, for \gpls\ of higher weights these relations are difficult to find starting from the integral
representation of the \gpls.
Consequently, relations of the type shown in Eq.~(\ref{eq:Gat1})
were found by means of numerical fitting routines working
with a precision up to $\sim 1000$ digits.

The \gpls\ of argument $\beta$ were then expanded in the limit $\beta \to 0$. The
expansions of the individual \gpls\ involve powers of $\beta$ and $\ln \beta$, HPLs of
argument $\xi$ and weights $\{0,1\}$, as well as the constants mentioned above. When
the expansions of the individual \gpls\ are inserted in the color coefficients,
 the $\ln \beta$ as well as the HPLs
of argument $\xi$ cancel out. The same cancellation was observed already in the
analogous expansion of the result in \cite{Bonciani:2010mn}. The expanded results
depend only on powers of $\beta$ (starting, in the case of the color coefficients $F_l$
and $G_l$, from the power $-1$, which characterizes Coulomb singularities) and on
powers of $\Xxi$. The forward-backward symmetry of the result is
manifest. It is also striking to observe that the many irrational constants appearing
in the expansion of the individual \gpls\ combine in such a way that in the expansion of
the color coefficients only the constants $\zeta_2$, $\zeta_3$ and $\ln 2$ are present.

\subsection{Small mass expansion}

In this Section we discuss the expansion of our results in the small-mass (or high-energy) limit, 
defined as the region where $s,|t|,|u| \gg m^2$. It
must be noted that the two-loop corrections to the top-quark pair production at leading
order in this limit were calculated a few years ago \cite{Czakon:2007ej,
Czakon:2007wk}. We can therefore test our calculation by comparing the leading order
of the expansions we present here with the results available in the literature.

The expansions are parametrized by the ratio $m^2/s \to 0$, where the
dimensionless parameter $\phi=-(t-m^2)/s$ is kept fixed. The expansions of the seven
color coefficients in the $m^2/s \to 0$ limit up to and including terms of ${\mathcal
O}((m^2/s)^5)$ can be found in the ancillary file \verb+ggtt-lightnf-smallmass.m+
included in the arXiv submission of this work. The leading terms in these expansions
fully agree with the corresponding results in \cite{Czakon:2007wk}, provided that one
accounts for the difference in the  overall normalization factor. For this reason, we
do not type these results here. The right column in Fig.~\ref{fig:expansions} compares
the exact results for the color coefficients with the small mass expansions obtained
by retaining terms of ${\mathcal O}((m^2/s)^0)$, ${\mathcal O}(m^2/s)$
and ${\mathcal O}((m^2/s)^5)$. In those plots we set $\phi=0.35$ and
$\mu=m$.  As expected, for large values of $\eta$, all of the expansions reproduce the
exact result.

In order to obtain expansions in the small mass limit, it is convenient to first
rewrite the color coefficients in terms of \gpls\ of arguments $\{x,\phi\}$. (Note
that the more obvious choice $\{m^2/s,\phi\}$ requires the introduction of square root
factors which make subsequent algebraic manipulations inconvenient.) The color
coefficient can be written in terms of \gpls\ of argument $x$ and weights belonging to the set
\begin{displaymath}
\left\{-1, 0, 1, -\frac{1 - \phi}{\phi}, -\frac{\phi}{1 - \phi}, [1 + o^2], 
[1 - o + o^2], \left[1 + o \frac{1-2\phi}{\phi}  + o^2\right],
\left[1  - o \frac{1-2 \phi}{1-\phi} + o^2\right]\right\} ,
\end{displaymath}
together with HPLs or argument $\phi$ and weights $\{0,1\}$. The only transcendental
constants involved are powers of $\pi$ and $\zeta_3$. The expansion of the \gpls\ of
argument $x$ can be carried out by first expanding them in the $x \to 0$ limit and 
then by expanding $x$ in powers of $m^2/s$:
\be
x = -\frac{m^2}{s} - 2 \left(\frac{m^2}{s}\right)^2 - 5 \left(\frac{m^2}{s}\right)^3 
- 14 \left(\frac{m^2}{s}\right)^4+ \cdots \, .
\ee
The expansion of these \gpls\ in $m^2/s$ involves logarithms of the same ratio,
HPLs of argument $\phi$ and
irrational constants which depend on $\zeta_3$ and powers of $\pi$. We insert
the expansions of the \gpls\ of argument $x$ in the color coefficients
and rewrite the HPLs of argument $\phi$ and weights $\{0,1\}$
in yet another form.
Here, it is sufficient to choose logarithms and $\Li_n$ ($n=2,3,4$) of
arguments $\phi$  and $1-\phi$, as well as 
\begin{displaymath}
\Li_4 \left(-\frac{\phi}{1-\phi}\right) \, , \qquad 
\Li_4 \left(-\frac{1-\phi}{\phi}\right) \, .
\end{displaymath}
In order to have the invariance of the result under 
the transformation $\phi \to 1-\phi$ (forward-backward symmetry)
manifest, we arrange for a functional basis such that the functions listed
above enter in the combinations ($f(\phi)+f(1-\phi)$) or ($(2\phi-1)(f(\phi)-f(1-\phi))$).
The cofactors of these building blocks depend on $\phi$ only through powers of
$(2\phi-1)^2 - 1$; spurious poles at $\phi = 1/2$ which appear at intermediate
stages cancel out.
At this point, the final results for the expansions are manifestly forward-backward
symmetric and numerically stable.


\section{Conclusions \label{sec:concl}}

In this paper we obtained analytic expressions for all of the two-loop corrections
with a closed light-quark loop which contribute to top-quark pair production in the
gluon-fusion channel. The calculation is carried out by retaining the full dependence
on the top-quark mass. The two-loop diagrams are interfered with the tree-level diagrams, 
and the results are presented as analytic formulas for the corresponding seven
gauge-independent color coefficients in the NNLO virtual contributions proportional
to the number of light quarks $N_l$ (see Eq.~(\ref{eq:cocoef})). The analytic results are
collected in ancillary files included in the arXiv submission of this work. For
illustrative purposes, we present plots of the
finite parts of the color coefficients as functions of the dimensionless parameters
$\eta$ and $\phi$, and we provide numerical values at three
different benchmark points in the physical phase space.
Furthermore, in this work it was shown 
that, after the results are written in a suitable form, expansions in the production
threshold and small mass limits can be obtained in a straightforward manner. The
expanded color coefficients in both limits are also provided in ancillary files
included in the arXiv submission of this work.

In comparison to analytic results for the two-loop corrections involving a
closed quark loop in the quark annihilation channel \cite{Bonciani:2008az} and
for corrections contributing to the leading color coefficient in both channels
\cite{Bonciani:2009nb,Bonciani:2010mn}, the results discussed in the present
work have a more involved structure. The  number of \gpls\ and the kind of
weights which arise in the calculation of the box MIs
\cite{vonManteuffel:2013uoa} made it mandatory to devote a large amount of work
to the efficient organization and  simplification of the final result. This
required numerous non-trivial rewritings of the color coefficients using
different sets of functional bases depending on different pairs of dimensionless
variables, as well as a mapping of the \gpls\ to real valued functions
specifically suitable for fast and stable numerical evaluations.

In particular, we show that the exact results for the color coefficients can be
expressed in terms of logarithms, classical polylogarithms $\Li_n$ ($n=2,3,4$),
and  genuine multiple polylogarithms $\Li_{2,2}$. Here, the functional arguments
are chosen such that the functions are real valued over the entire physical
phase space. This change in the functional basis was made possible by an
extension of the algorithms given in \cite{Duhr:2011zq,Duhr:2012fh} suitable to
handle \gpls\ depending on the generalized weights employed for non--linear
denominators in the integrating factors in this work. For the first time, we
demonstrated that such a basis choice is possible also in the presence of
generalized weights for a multiscale QCD application. We emphasize that no
spurious symbol letters needed to be introduced and all generalized
weights could be eliminated. We observe that the numerical evaluation of the color
coefficients is faster by an order of magnitude when this optimized functional
basis is employed.

It will be interesting to see if the methods developed and employed for the
results presented here can be applied to the calculation of the subleading
two-loop color coefficients in the quark-annihilation channel. These subleading color
coefficients involve a number of non-planar diagrams with massive internal and external
particles and to date they were calculated only numerically \cite{Czakon:2008zk}.
In the gluon-fusion channel, the subleading color coefficients $B$, $C$, $D$ and
the heavy-quark contributions $E_h$, $F_h$, $G_h$ are known only
numerically \cite{BaernreutherTh}. The analytic calculation of these gluon
initiated corrections poses an additional independent challenge, since they
are known to depend on MIs which involve elliptic integrals.
Those structures arise from diagrams which include a sunrise subtopology with three
massive internal legs and a momentum transferred which is not on the mass shell of
the internal legs \cite{Laporta:2004rb}. Recent developments
\cite{MullerStach:2011ru,Adams:2013nia} give
rise to the hope that a better understanding of
the analytic structure of massive Feynman diagrams will also allow for an analytic
evaluation of these sets of corrections.


\acknowledgments{
Part of the algebraic manipulations required in this work were carried out with {\tt
FORM} \cite{Kuipers:2012rf}. The Feynman diagrams were drawn with {\tt Axodraw}
\cite{axodraw}. The work of AvM\ was supported in part by the Schweizer
Nationalfonds (Grant 200020\_124773/1), by the Research Center {\em Elementary Forces
and Mathematical Foundations (EMG)} of the Johannes Gutenberg University of Mainz and
by the German Research Foundation (DFG).  The work of AF\ was supported in part by
the PSC-CUNY Award No.\ 66590-00-44 and by the National Science Foundation Grant No.\
PHY-1068317. The work of RB\ was supported by European Community Seventh Framework
Programme FP7/2007-2013, under grant agreement N.302997.
}


\appendix

\section{Numerical samples}
\label{sec:numbers}

In this appendix, we present numerical results for all of the seven color
coefficients we calculated. Each one of the following tables shows values of the
pole coefficients and finite parts for a different benchmark point in the
physical phase space. The results shown here are normalized in the standard 
$\overline{\mbox{{\small{MS}}}}$ way, that is a factor $((4\pi)^\varepsilon
e^{-\gamma \varepsilon})^2$ is extracted and not included in the numerics. That
means that the results in the ancillary files are first multiplied by a factor
$e^{2\gamma \varepsilon} \Gamma^2(1+\varepsilon)$ and then expanded in 
$\varepsilon$ before the numbers are obtained.

For $s/m^2 = 5$, $ t/m^2 = -1.25$, $ \mu/m = 1$ we obtain

\begin{center}
\begin{tabular}{|c||c|c|c|c|}
\hline Coeff.  &$1/\varepsilon^3$  & $1/\varepsilon^2$ & $1/\varepsilon$ & $\varepsilon^0$ \\
\hline 
\hline$E_l$  &$-0.7838122811$ & 1.137904118 & 1.747317683 &$-7.037881174  $ \\ 
\hline $F_l$ & 1.552103527 & $-1.663042888$ & $-3.172113037$ & 7.815862217    \\ 
\hline $G_l$ & -  & 0.1937203624 & 4.190187785  &  $-13.38175914$   \\ 
\hline  $H_l$& - & 0.1492975773 & $-0.3407519641$  &  0.2270538721  \\ 
\hline  $H_{lh}$& - &-  & $-0.0002689061861$ &   $-0.2466070280$ \\ 
\hline  $I_l$& - & $-0.2956387670$ & 0.6756453423  & $-0.4489337823$  \\ 
\hline  $I_{lh}$& - &-  & -  &   0.4863062794 \\ 
\hline 
\end{tabular}
\end{center}

For $s/m^2 = 43$, $ t/m^2 = -21$, $\mu/m = 1.7$ we obtain

\begin{center}
\begin{tabular}{|c||c|c|c|c|}
\hline Coeff.  &$1/\varepsilon^3$  & $1/\varepsilon^2$ & $1/\varepsilon$ & $\varepsilon^0$ \\
\hline 
\hline$E_l$  &$ -0.6828131172$& 2.465864806& $-2.816012729$ &   $-2.086232939$ \\ 
\hline $F_l$ & 1.364888058 &$ -4.939937930$ & 4.260219926 & 11.41745753   \\ 
\hline $G_l$ & -  & 1.589549879  &  $- 3.010243755$ &   $-10.88329499$ \\ 
\hline  $H_l$& - & 0.1300596414 & $- 0.3337514448$  &  0.2224382692  \\ 
\hline  $H_{lh}$& - &-  & $- 0.2760509183$ &  0.3479512918 \\ 
\hline  $I_l$& - & $- 0.2599786777$ & 0.6671476671 &  $-0.4446849447$  \\ 
\hline  $I_{lh}$& - &-  & 0.5518081243 & $-0.6955768357$  \\ 
\hline 
\end{tabular} 
\end{center}

For $s/m^2 = 8.1$, $ t/m^2 = -0.6$, $\mu/m = 2.1$ we obtain

\begin{center}
\begin{tabular}{|c||c|c|c|c|}
\hline Coeff.  &$1/\varepsilon^3$  & $1/\varepsilon^2$ & $1/\varepsilon$ & $\varepsilon^0$ \\
\hline 
\hline  $E_l$    & $-1.991067909$ & $- 0.1771655934$ & 12.03682866&  $-24.21746110$  \\ 
\hline  $F_l$    & 2.915285996 & $- 0.07400739283$ & $-17.47415387$ & 35.65735116    \\ 
\hline  $G_l$    & - & 1.105009458 & 3.962101090 &   $-15.96447616$ \\ 
\hline  $H_l$    & - & 0.3792510302 & $- 0.7898719870$  & 0.4900775204   \\ 
\hline  $H_{lh}$ & - &-  & $-1.131777182$ & 0.8530176592 \\ 
\hline  $I_l$    & - & $- 0.5552925707$ & 1.179700855&   $-0.7009615385$  \\ 
\hline  $I_{lh}$ & - &-  & 1.647969182 & $- 1.364946933$ \\ 
\hline 
\end{tabular} 
\end{center}


\end{document}